\documentclass[sigconf]{acmart}

\usepackage{amsmath,amsfonts}
\usepackage{algorithmic}
\usepackage{graphicx}
\usepackage{textcomp}
\usepackage{xcolor}
\usepackage{verbatim}

\def\BibTeX{{\rm B\kern-.05em{\sc i\kern-.025em b}\kern-.08em
    T\kern-.1667em\lower.7ex\hbox{E}\kern-.125emX}}
\begin{document}


\title{Changing Software Engineers' Self-Efficacy with Bootcamps: A Research Proposal}

\author{Danilo Monteiro Ribeiro}
\authornote{Both authors contributed equally to this research.}
\email{danilo.ribeiro@zup.com.br}
\affiliation{%
  \institution{ZUP Innovation}
  \city{São Paulo}
  \state{São Paulo}
  \country{Brazil}
}

\author{Alberto Souza}
\email{alberto.tavares@zup.com.br}
\affiliation{%
  \institution{ZUP Innovation}
  \city{São Paulo}
  \state{São Paulo}
  \country{Brazil}
  }

\author{Victor Santiago}
\email{victor.pinto@zup.com.br}
\affiliation{%
  \institution{ZUP Innovation}
  \city{São Paulo}
  \state{São Paulo}
  \country{Brazil}
  } 

\author{Danilo Lucena}
\affiliation{%
  \institution{ZUP Innovation}
  \city{São Paulo}
  \state{São Paulo}
  \country{Brazil}
  }
\email{danilo.lucena@zup.com.br}

\author{Geraldo Gomes}
\email{geraldo.junior@zup.com.br}
\affiliation{%
  \institution{ZUP Innovation}
  \city{São Paulo}
  \state{São Paulo}
  \country{Brazil}
  }

\author{Gustavo Pinto}
\email{gustavo.pinto@zup.com.br}
\affiliation{%
  \institution{ZUP Innovation}
  \city{São Paulo}
  \state{São Paulo}
  \country{Brazil}
  }

\begin{abstract}In several areas of knowledge, self-efficacy is related to the performance of individuals, including in Software Engineering. However, it is not clear how self-efficacy can be modified in training conducted by the industry. Furthermore, we still do not understand how self-efficacy can impact an individual's team and career in the industry. This lack of understanding can negatively impact how companies and individuals perceive the importance of self-efficacy in the field.
Therefore, We present a research proposal that aims to understand the relationship between self-efficacy and training in Software Engineering. Moreover, we look to understand the role of self-efficacy at Software Development industry.  We propose a longitudinal case study with software engineers at Zup Innovation that participating of our bootcamp training. We expect to collect data to support our assumptions that self-efficacy can be related to training in Software Engineering. The other assumption is that self-efficacy at the beginning of training is higher than the middle, and that self-efficacy at the end of training is higher than the middle. We expect that the study proposed in this article will motivate a discussion about self-efficacy and the importance of training employers in the industry of software development.
\end{abstract}
\maketitle


\section{Introduction}

Self-efficacy as a belief in one's capabilities to organize and execute the courses of action required to produce a given attainment~\cite{bandura1977self}. More concretely, self-efficacy is how much an individual~\emph{believes} that s/he is capable of performing an specific task. For example, when a developer assesses that s/he is able to finish all her/him task by the end of the week, based on her/him experience in conducting similar tasks, this developer is exercising her self-efficacy.
Self-efficacy determines the effort that the individual expended to cope with a task, even if it has impediments and difficulties~\cite{bandura1977self}.
An individual who has low self-efficacy tends to avoid difficult tasks and put less effort into completing a task~\cite{artino2012academic}.For example, if a developer believes he is not able to act on a certain part of the code, he may not pick up activities related to that part. Or, if he doesn't believe he's able to learn some technology in time to apply it to the project, he may be less effortful about doing this task.

Software engineers are also subject to the self-efficacy effect. For instance, it was observed that individuals may have high or low self-efficacy \cite{hazzan2010recruiting}. 
If software engineers have low self-efficacy but high skills, their performance could be compromised because they might not believe in their potential ability to perform certain tasks and thus they can strive less~\cite{bandura2010self}, which could impact their professional development and career progress. For instance, engineers may not apply for more senior positions because they do not feel they are capable for performing the duties.
On the other hand, if engineers have high self-efficacy but low skills, their performance can be compromised. For example, the weekly sprint could be mistakenly estimated, adding more tasks that they are able to conduct during the sprint~\cite{vancouver2002two,pajares1996self,schunk2009self}

Some authors have been investigating the students perception about their self-efficacy. 
Moores and Chang~\cite{moores2009self} conducted a study with software engineering students and found that overconfidence in self-efficacy is negatively related to performance. The authors also found that when there was an intervention (for example, if a mentor gives a task performing feedback), self-efficacy was positively related to performance. Another study found that algorithms self-efficacy can be improve by pedagogic interventions \cite{toma2018self}.
One important characteristics of self-efficacy is that one could calibrate it. For example, Brannick et al. \cite{brannick2005calibration} found that mentoring can calibrate self-efficacy of psychology's students. 

For instance, by know that an engineer has low self-efficacy, this engineer could use some practices and tools to increase it. Although there are a couple of studies that investigate self-efficacy in a classroom setting, little is know about the role of self-efficacy in modern software development bootcamps.

Coding bootcamps (or bootcamps for short) can be defined as a training program focused on teaching programming-related concepts that are demanded by software producing companies to developers in the initial levels of their software development career~\cite{waguespack2018triangulating}. 

In other areas, such as health, bootcamps are already used to develop more qualified professionals with better experiences to work on a daily basis\cite{esterl2006senior}.
Coding bootcamps are, however, becoming commonplace in the software development industry.
Developers participate in bootcamps due to several reasons, including: they believe that bootcamp focus on the most important aspects of coding, the application of learned knowledge to solve real-world problems to and the potential opportunity to find a job in a tech company~\cite{firehouse}. 
In a survey with 105 bootcamps and 133k bootcamp graduates, it was noticed that a $\sim$5\% in the number of students participating in bootcamps (from 2018 to 2019); in the same period, however, online bootcamps grew 31.44\% \cite{karma}. As a consequence, bootcamps have become an important source for companies looking to hire technical talents. Some organizations have developed their own bootcamps seeking to improve their internal workforce~\cite{navas}.




The goal of this paper is to present a proposal to understand the role of self-efficacy in coding bootcamps. We present and evaluate an approach to assess and calibrate self-efficacy in bootcamps. This bootcamps were organized by a  Brazilian tech company, with about 3k employees.


\section{Background}
\label{background}

\subsection{Self-efficacy Theory}

Self-efficacy is proposed as a construct of Social Cognitive Theory~\cite{bandura1977social} and can be defined as the belief in one's own ability to accomplish something successfully~\cite{bandura1977self}.

Self-efficacy can change the perception of reality and how individuals behave~\cite{bandura2010self}, the main assumption of self-efficacy is that people generally will only attempt things they believe they can accomplish and will not attempt things they believe they will fail~\cite{bandura1977self}. This is because individuals generally do not put effort to do a task that they believe they cannot do~\cite{bandura1977self}. When individuals have a strong sense of efficacy, they trust that they can accomplish even the most difficult tasks. Furthermore, they can see the task as challenges to be mastered, rather than threats to be avoided~\cite{bandura1994ramachaudran}.

Bandura affirms that there are four sources of self-efficacy.  The first and most effective way to impact self-efficacy is by mastering experiences~\cite{bandura1994ramachaudran}. The process of mastering an experience is based on doing a task and succeeding in doing it. This occurs because individuals tend to believe that they are more capable to do a task if they have performed well a similar task before.

A famous example of mastery experience is babysitting. Women who have experience taking care of kids before becoming mothers are more confident with their abilities to take care their kids~\cite{froman1989infant}.
Therefore, to master at something, an individual has to practice the task. However, Bandura~\cite{bandura1994ramachaudran} presents some important points. For example, if the tasks are easy or very similar to the ones already performed, self-efficacy may be poorly developed by individuals. Moreover, the task needs to be difficult but not impossible. The individual needs to approach difficulty tasks and work through obstacles to improve self-efficacy~\cite{bandura2010self}. 
Heslin~\cite{heslin1999boosting} gives another tip to improve mastery experience: breaking down a difficult task into small steps, which are relatively easy, to ensure a high level of initial success, then progressively increase the difficulty of the task so that the participant still feels challenged. At last, Heslin suggests providing feedback through workshops, training programs, internships, and clinical experiences to improve mastery experiences. 

Another factor is vicarious experience. Vicarious experience can occur when an individual observed others' successes and failures who are similar to him/her. Bandura~\cite{bandura1977self} mentioned that watching someone like you (e.g., a colleague in a similar level of experience) doing and successfully accomplishing something you would like to do (e.g., a task during the spring) can increase \emph{your} self-efficacy. Conversely, when you are observing someone like you to fail, \emph{your} self-efficacy tends to be negatively impacted. However, failure is not always harmful; when they feel confident, individuals can avoid repeating the errors they observed others doing~\cite{brown2013self}. 
Heslin~\cite{heslin1999boosting} commented that to improve self-efficacy, a manager can do workshops and training sessions because when individuals are watching others in a training session, the manager can provide observational experiences.
 
The third factor affecting self-efficacy is verbal persuasion. When individuals are persuaded verbally, they may become more confident that they can do the task~\cite{bandura1977self}. 
Having others verbally supporting the attainment of a task helps to support a person's belief in himself or herself~\cite{bandura1977self}. 
Verbal persuasion is frequently used by sports coaches to improve self-efficacy~\cite{brown2013self}. Verbal persuasion builds self-efficacy when managers encourage and praise their competence and ability to improve their performance~\cite{heslin1999boosting}. 
 
The fourth and last factor is physical and emotional states that occur when someone contemplates doing something~\cite{bandura1977self}. For example, anxiety, fear, and worry can affect self-efficacy negatively and can lead an individual to believe that she cannot perform a task~\cite{pajares1996self}.

\subsection{Self-efficacy in Software Engineering}

Some studies investigated self-efficacy  in Software Engineering. Tsai and Cheng~\cite{tsai2010programmer} conducted an industrial research that found self-efficacy is related positively to intention to knowledge sharing and knowledge sharing. The authors hypothesized that self-efficacy could improve knowledge sharing at software teams.
Another study investigated that the students' programming self-efficacy beliefs had a strong positive impact on the effort and persistence of Software Engineering students. Furthermore, it was found that self-efficacy is negatively related to seeking help~\cite{kanaparan2017self}. Arya et al.\cite{arya2012moderating} found that self-efficacy is positively but not very significantly related to organizational commitment. Fu~\cite{fu2010information} found that professional self-efficacy is positively related to the career commitment of Software Engineers. 
Hazzan and Seger~\cite{hazzan2010recruiting} identified in their study that high self-efficacy practitioners tend to be: "\textit{more cooperative, have a greater sense of morale working with their team members, feel that their relationships with co-workers are closer, get better managerial support, report higher needs in achievement, dominance, affiliation, and difference and have better attitudes towards change}". 

Other studies are aiming to understand how to improve self-efficacy. For example, Dunlap~\cite{dunlap2005problem} observed that the use of Problem-Based Learning could improve the self-efficacy of Software Engineering students. Steinmacher et al.~\cite{steinmacher2015increasing} found that an online coach called FLOSScoach had a positive influence on open-source newcomers' self-efficacy, making newcomers more confident and comfortable during the project contribution process.
Srisupawong et al.~\cite{srisupawong2018relationship} revealed that perceptions of autonomy, meaningfulness, and involvement are positively associate with strong self-efficacy. Furthermore, the students' perceptions of vicarious experiences and perceptions of social persuasions demonstrated a positive relationship with self-efficacy. Perceived physiological and affective states demonstrated a negative influence self-efficacy of computer science students.

\subsection{Bootcamps}

Early uses of bootcamp dates from the intensive physical training that sought to make prisoners physically active and encourage discipline in the 19th century~\cite{anderson1999boot}. Later, in training camps, American sailors received a set of intense activities during the American-Spanish war~\cite{bootcampHistory}.

Bootcamps are also widely used in health education. The objective is to accelerate the learning process about a given topic following a practical and dynamic way~\cite{esterl2006senior}.
In general, students have little or no experience previous experience in the subject covered. From the bootcamp onwards, they are expected to master the skills studied, helping them to mitigate eventual skills gaps.

Some studies investigated the relationship between bootcamp and self-efficacy. Smith et al~\cite{smith2019condensed} investigated students in their final year of an entry-to-practice master of occupational therapy program. They found that students’ self-efficacy for assessing, training, spotting, and documenting manual and power wheelchair skills improved by between 28\% and 35\%.
Devon et al. developed a pediatric bootcamp where medical students participated in practical sessions on their day-to-day work. They observed that the subjects' self-efficacy increased in all investigated items~\cite{devon2019pediatric}.

\section{Our Bootcamp context}\label{sec:bootcamps}

Our organization has intense training as one of its main pillars.  We believe that focus on the training process should be part of every developer's learning journey.
Our bootcamp can be seen as a career acceleration program and seeks to improve one's path in technology, through an intense and deliberate training perspective. During the last year, we conducted a total of 11 bootcamps.
The goal of the bootcamp is to prepare the bootcamp participants (and future company's employees) to situations that may occur in the teams to which they will be allocated, arming them with knowledge and tools to solve the most common problems. Thus, our bootcamp training is based on real situations that are needed to perform daily coding activities in our organization.  Our bootcamp takes from 12 to 16 weeks, and have between 10 and 80 participants (depends on the needs of the organization).
Next we further detail our bootcamp protocol.


\subsection{Bootcamp participants}

We create open calls for junior developers all around the country. Although the needed skills vary from bootcamp to bootcamp, overall anyone with basic programming skills (e.g., Java or Kotlin programming; Hibernate and SQL for managing databases; Spring MVC for structuring web applications; Git for distributed code sharing) could apply to participate in the bootcamp. We also require that the participants should be available to work 40h peer week. During the 11 bootcamps we conducted, we received more than 12k applications, where around 450 participants participated in our bootcamp. Most of these participants are junior developers (some with no previous professional experience).



\subsection{Bootcamp mentors}
A team of experts in the technologies used in the company was recruited to create our bootcamp protocol. These technical mentors (so called mentors) were also responsible for teaching, mentoring and evaluating the participants. 
We have two mentors for each bootcamp. Mentors are also well-known figures in the tech radar in Brazil, including renowned speakers and event organizers related to these technologies. Some even have books among the bestsellers about the technology used in the bootcamp. 
All mentors are full time dedicated to bootcamp execution. Besides from the technical mentors, the bootcamp participants are also closely followed by a program manager. The program manager monitors the development of training participants, ensuring the best experience during the program, supporting the mentor team.

\subsection{Bootcamp sessions}
The bootcamp had synchronous and asynchronous sessions. The training occurs mostly on synchronous sessions. 
The mentors give one hour of synchronous theoretical class per day and two more synchronous sessions of two hours each during the week. At those sessions, the mentors given feedback and answer questions. For example, prior to the session, the mentors look at the participants' source code, raise points for improvements, and highlight good coding patterns. Mentors showcase these code snippets during the sessions to the participants so that they can discuss and understand the rationale behind the code changes. The mentors are available to clarify any doubts during working hours, in a virtual room. 
There are also coding sessions such as dojo and pair programming that take place during the training. To help participants further explore their learning path, we also provide to them several asynchronous programming courses. These courses are made by a third-party company, which the company bought a license that allow the participants to access the courses.
The bootcamp approaches several technical subjects, including: 
Software development (Java, Kotlin, Spring Boot, SQL and Micronaut), Software design (SOLID), Software architecture (Distributed Systems and HTTP),  Software testing and Software Observability (Prometheus, Grafana, RED and USE metrics). All topics were chosen due to the company needs. The full list of technical subjects is available online at: \texttt{https://bit.ly/3a9I7cB} 

\subsection{Bootcamp evaluation}

At the beginning of each week, we plan to deliver a self-efficacy questionnaire to the bootcamp participants; since our bootcamps have on average 12 week, we expect to deliver 12 self-efficacy questionnaires per bootcamp. The goal of these self-efficacy questionnaires is to assess how much the participants believe they are capable to perform the software development activities. We provide more information about these self-efficacy questionnaires in Section~\ref{sec:questionnaires}.

For each technical subject, there is a set of key items raised by the experts that the participant should be able to perform at the end of the course. These items are on a scale from 0 to 10 (where 0 means Not to all confident and 10 means Extremely confident). The participant will answers a questionnaire (example\footnote{https://encurtador.com.br/noPZ6}) with these items at the beginning and at the end of each technical subject. In addition, at the end of the technical subject, the participant also has a set of exercises about it. 

After answering the exercises, the participants can compare their responses with the mentors' responses; The participants answers are stored in a database.
Therefore, when all participants finish a topic, mentors have access to the participant's responses and prepare feedback sessions with the most common mistakes, as well as examples of responses that were well constructed.

The mentors also perform 1:1 feedback with participants, where they use the responses sent to question situations, praise responses, and propose improvements.


\section{Research Questions}

The goal of this study is to understand the role of self-efficacy in the coding bootcamps. We plan to conduct this investigation in a software producing company in Brazil, the same one that performs the bootcamps. 

\subsection{Research Questions}

In this work, we plan to investigate two research question.
First, we want to understand how software engineer's self-efficacy and the lack of self-efficacy calibration can impact their work. 
Therefore, we pose the following research question (RQ):

~\textbf{\textit{RQ1: How self-efficacy can impact the career and teamwork of software engineer who participated in an industrial bootcamp? }}


Next, we want to understand how the participants perceive their self-efficacy being impacted during the training. Our hypothesis is that: at the beginning of the training, participants' self-efficacy may be uncalibrated with their performance. For instance, since our bootcamps focused on training junior software engineers, these participants are more likely to have little knowledge in the software development stack we trained them on. 
We believe that as new skills are introduced during the training process, the participants might have less perception of their self-efficacy. However, when participants challenge themselves to perform tasks, and receive good feedback they might be more calibrate their self-efficacy with their real skills. 

It is important to note that the (re-)calibration process can be more challenging in the online bootcamp scenario because participants have less contact with their mentors, and therefore they have to assess their own self-efficacy. The goal of our next research question is to provide means to help mentors to assess the participant's self-efficacy without have constant contact with them. 

Thus, our second research question is the following:

\textbf{\textit{RQ2: How to calibrate software engineers' self-efficacy in a industrial bootcamp?}} 

By answering these questions, we hope we could improve software engineering training, in particular, in the context of bootcamps, which are becoming even more prevalent in the software development industry. As a possible outcome of this research, 
we plan make recommendations for future industrial training and academic courses alike about approaches for measuring self-efficacy--- and how we could better calibrate it. 

Moreover, by exploring this gap, we will make recommendations for future academic courses and industrial training about measuring self-efficacy and how we can calibrate it. 
Finally, we want to propose a structure to monitor and measure the self-efficacy of individuals in bootcamps carried out by the industry.

\section{Methodological Framework} 

This investigation will be conducted in the context of a large (3k+ employees) Brazilian software producing company: Zup Innovation. Itaú Unibanco recently purchased Zup innovation. Itaú Unibanco is the largest private bank in Brazil, the largest financial institution in Latin America, and one of the largest in the world\footnote{https://www.itau.com.br/institucional/sobre-o-itau/}. 
The empirical method we plan to use for this proposal is the case study. A case study can be defined as an empirical inquiry that investigates a contemporary phenomenon within its real-life context~\cite{easterbrook2008selecting, runeson2012case}
We intend to use a prospective longitudinal design, when the participants are followed over a period of time. In our study, the participants will be evaluated during 1) the bootcamp and 2) at the beginning of their career at the company.

This research will collect qualitative and quantitative data at different times.
Runeson and Host~\cite{runeson2009guidelines} state that the researcher must clarify their research objectives and questions when planning a case study. The authors also state that the purpose of the case study may be to explore, describe or explain a given phenomenon.
In this research, the phenomenon is self-efficacy, and the objective is to understand how self-efficacy impacts software development. Moreover, understand how self-efficacy is affected by training in the industrial context. Thus, the case of the study is how self-efficacy works and shifts from bootcamps conducted by the software industry. 

In the view of Yin~\cite{yin2009case}, to develop a case study, it is also necessary to have a well-defined unit of analysis. In this work, two units of analysis will be used: The class in which the students will participate in the training and the individuals who will participate in the training. The class was chosen because we are going to group the results in the class that they are performed to understand the context that class is going through. Individuals were chosen because self-efficacy is an individual construct.

Given this context, our study can be characterized as a multiple case study with the embedded approach~\cite{yin2009case}. Yin~\cite{yin2009case} also comments that multiple case studies present more convincing evidence and more robust overall results, as single case studies may be limited by the unique conditions surrounding the case.

To conduct this research, we plan to conduct ten bootcamps, with approximately 50 participants per bootcamp. As the reader observed in Section~\ref{sec:bootcamps}, we have extensive experience in conducting bootcamps.
All subjects will pass through an on-line recruitment process design by bootcamp's organizers. The organization is aimed to hire all participants that successfully complete the bootcamp. It is possible that some participants may leave (or even be fired) during the process. For those that complete the bootcamp, they will be allocated to a software development team at the organization.

\subsection{Data Collection}\label{sec:questionnaires}

We plan to gather quantitative and qualitative data. The quantitative data will be based on a family of questionnaires. The qualitative data will come from observations and interviews with the participants to get a more in-depth understanding about self-efficacy in industry's training.

\subsubsection{Quantitative}
We plan to deploy a family of questionnaires.
First, we will gather preliminary information from online questionnaires, like participants' demographics, programming experience and education level, and then we will conduct semi-structured interviews to get a more in-depth discussion about self-efficacy and training in the industry.

Upon entering the training program, all participants should answer a quantitative self-efficacy questionnaire. This questionnaire was inspired by the one developed by Steinhorst et al.~\cite{steinhorst2020revisiting}. It approaches four dimensions of self-efficacy: (i) Tracing Program Flow, (ii) Using Structures and Patterns for Problem-Solving, (iii) Persistence, Debugging, and Problem-Solving Competences, and (iv) Controlling Program Flow. We remove the first dimension (Tracing Program Flow) because it was out of the scope of the organization goals to explore it.
We also plan to apply this questionnaire at the end of the bootcamp process to understand eventual changes in the participants' perception. This questionnaire is called as \emph{basic self-efficacy}.

In addition, we also designed another quantitative self-efficacy questionnaire focused on software developer roles. In this second questionnaire we  evaluate: (i) Performance (How much the individual feels able to perform activities and perform well), (ii) Technical Mastery (how much the individual believes he knows relating to his technical level), (iii) and Teamwork (how much that the individual believes he/she can contribute to teamwork and influence others to do their best).
Unlike Steinhorst's et al. questionnaire~\cite{steinhorst2020revisiting}, in this questionnaire we assess the skills and behaviors that the company perceive as valuable for software engineers. This questionnaire will be evaluated weekly to follow the evolution of the participants during the bootcamp. This questionnarie is called as \emph{Role guide self-efficacy}. The English version of this questionnaire can be seen here:  \texttt{https://encurtador.com.br/hFLX0}

Finally, for each subject taught during the bootcamp training, we will apply a self-efficacy assessment questionnaire about that subject. After completing this questionnaire, participants will have to answer an assessment developed and evaluated by the mentors. This questionnaire is called as \emph{course self-efficacy}  

\subsubsection{Qualitative}

Throughout the process, the mentors have to assess the students' ability to perform as a Software Engineer. This assessment can be made through pair programming and coding dojo sessions. These sessions are (video and audio) recorded. By the triangulation of these data, we intend to measure the real ability of individuals to compare with their self-efficacy and their calibration in different times of training. 

Finally, participants will go through semi-structured qualitative interviews during the bootcamp and after their allocation to software development teams to understand their perception of self-efficacy and its calibration.
This questionnaire can be view here~\footnote{https://bit.ly/3iNO5Vo}

The career success of individuals in the company does not depend on their participation in the questionnaire. Before collecting the data, the research team is introduced, clarifies what the research is about, and explains the importance of being honest in the answers to the questionnaires. The bootcamp participants signed a consent form in their first week. Each week, along with the questionnaire, a brief description of the research, the confidentiality of information, and the importance of honestly answering the questionnaire are sent. 

\section{Initial results}

At this section, we present some initial results. Since we are going to measure self-efficacy, it is very important to have a questionnaire that actually reflects the individuals' self-efficacy. To validate the questionnaire, psychometric techniques were used based on the recommendations proposed by Hair et al.~\cite{hair2009multivariate}. Although these techniques already exist and are used in Software Engineering, they are recently gaining popularity in the field for increasing the reliability of research~\cite{graziotinPsycho}.

The \emph{Role guide self-efficacy} questionnaire  used in the bootcamp was applied to 108 company employees, and an exploratory factor analysis was carried out. 
Kaisor-Meyer-Olkin (KMO) and Bartlett’s sphericity test were used to measure the sampling adequacy.  The KMO test indicates the proportion of variance in your variables that might be caused by underlying factors. Values close to 1.0 generally indicate that a factor analysis may be useful ~\cite{ibm, hair2009multivariate}. Bartlett's test of sphericity tests investigate if the correlation matrix is an identity matrix, which would indicate that your variables are unrelated. Values than 0.05 indicate that a factor analysis may be useful ~\cite{ibm, hair2009multivariate}.

The results showed that the KMO value was 0.947, and the significance of Bartlett’s sphericity was 0.000. (v2 = 2915,422, df = 1,128, $\rho$ = 0.000), indicating that the good samples adequacy. 

After, principal component factor analysis was performed using Oblimax rotation with Kaiser Normalization. The main objective is analyzing the items to reduce a large dimension of data to a relatively smaller number of dimension. Here, we are looking to found the same three factors previously theoretically proposed for the questionnaire. (Performance, Technical Mastery and Teamwork). Factor analysis yielded a 3-factor solution with an explained variance of 73.81\%. 

Furthermore, We also used Cronbach's alpha to measure our scale reliability. We found $\beta$ = 0.953 at performance dimension, we found $\beta$ = 0.957 at domain dimension and for culture dimension, we found $\beta$ = 0.858. According to Streiner~\cite{streiner2003being}, values above 0.7 are considered with good reliability. All these results indicate a good quality of the scales  created in our bootcamp. 
In future works, we intend to present the whole process of scale's validation with more detail. Moreover, to increase the validity of the questionnaire, we intend to apply Confirmatory factor analysis to our scale too.

\section{Limitations and threats to validity}

Similar to all research, there are limitations to how this study was designed. First, we choose a case study and all investigation will be conducted on Zup Innovation. The context can limit some results. 

Regarding the participants, all of them will undergo a selection process led by human resources. We can specify attributes, but all selection of individuals is done outside the search. Therefore, we cannot guarantee that any aspect about the characteristics of the selected people or even that the selection process will be the same for all classes/participants.

Another important point is that the questionnaires are self-perceived, therefore, the honesty of the participants in their answers is necessary. We make clear the need to be honest with the questionnaire in the authorization and consent form. In addition, we held a presentation for participants explaining what research is all about. It is noteworthy that the organization already hires all people.

\section{Conclusions}

As mentioned above, self-efficacy can be related to performance and can be impacted by training. Therefore, this work aims to clarify the importance of developing self-efficacy with training in Software Engineering to the industry. 
The main contribution of this idea study proposed will be the identification of the relationship between training (bootcamp) and self-efficacy. Moreover, we pretend to understand how self-efficacy change during the weeks of training and the impact of self-efficacy in software engineering.  We believe that organizations should train and develop individuals' skills and their self-efficacy to perform better. 

Other relationships can be investigated in future works. For example, the relationship between self-efficacy and performance in an industrial context.  This relationship already was investigated in academic setting~\cite{davazdahemami2018training} in Software Engineering and is critical. 

Finally, this research should be an important step towards the consolidation of self-efficacy and training in industrial context in Software Engineering research.

\bibliographystyle{ACM-Reference-Format}
\bibliography{references.bib}

\end{document}